\documentclass{article}
\usepackage{spconf,amsmath,graphicx}
\usepackage[justification=centering]{caption}

\title{EACeleb: An East Asian Language Speaking Celebrity Dataset for Speaker Recognition}
\name{Desmond Caulley, Yufeng Yang, David V. Anderson \thanks{dataset and code is available at https://github.com/dcaulley/av\_diarization}}
\address{Georgia Institute of Technology}
\begin{document}
%
\maketitle
\begin{abstract}
Large datasets are very useful for training speaker recognition systems, and various research groups have constructed several over the years. Voxceleb is a large dataset for speaker recognition that is extracted from Youtube videos. This paper presents an audio-visual method for acquiring audio data from Youtube given the speaker's name as input. The system follows a pipeline similar to that of the Voxceleb data acquisition method. However, our work focuses on fast data acquisition by using face-tracking in subsequent frames once a face has been detected---this is preferable over face detection for every frame considering its computational cost. We show that applying audio diarization to our data after acquiring it can yield equal error rates comparable to Voxceleb. A secondary set of experiments showed that we could further decrease the error rate by fine-tuning a pre-trained x-vector system with the acquired data. Like Voxceleb, the work here focuses primarily on developing audio for celebrities. However, unlike Voxceleb, our target audio data is from celebrities in East Asian countries. Finally, we set up a speaker verification task to evaluate the accuracy of our acquired data. After diarization and fine-tuning, we achieved an equal error rate of approximately 4\% across our entire dataset.

\end{abstract}
\begin{keywords}
speaker verification, speaker diarization, speaker acquisition
\end{keywords}
\section{Introduction}
\label{sec:intro}

Researchers in the speaker recognition community has worked to build a database of audio of speakers in unconstrained or 'wild' conditions to develop more robust systems. One attempt at creating such a database is the Speakers in the Wild (SITW)\cite{sitw} speaker recognition challenge for Interspeech 2016. The database comprises hand-annotated speech samples from about 300 individuals. These audio recordings are unconstrained with recording locations from red carpet interviews, sports arenas, and the outdoors. Devices used to record the audio range from cellphones to camcorders. With such a dataset, researchers in the speech community can embark on the challenging task of building technology that is robust to real-world applications. The issue, however, is the scalability of building a hand-labeled dataset.

\begin{table}[h!]
    \centering
    \begin{tabular}{| c | c | c | c |}
         \hline
         \textbf{Dataset} & \textbf{Condition} & \# \textbf{POI} & \# \textbf{Utt}\\
         \hline
         ELSDSR &  Clean Speech & 22 & 198 \\
         \hline
         SITW & Multi-media & 299 & 2,800 \\ 
         \hline
         Forensic Comp.n & Telephony & 552  & 1,264 \\
         \hline
         Voxceleb 1 & Multi-media & 1,251 & 153,516 \\
         \hline
        \textbf{EACeleb} & \textbf{Multi-media} & \textbf{1,641} & \textbf{76,522} \\
         \hline
         Voxceleb 2 & Multi-media & 6,112 &  1,128,246 \\
         \hline

    \end{tabular}
    \caption{Comparison Unconstrained Datasets}
    \label{tab:ratio_data}
\end{table}

Voxceleb addressed this scalability issue by developing a fully automated pipeline that acquires celebrities' Youtube audio/video clips. Voxceleb queries Youtube by appending the word "interview" with the name of a celebrity and noting time segments in a Youtube video where the person of interest both shows up and speaks. With their system. Voxceleb acquired audio for over 7000 speakers with at least 2000 hours of audio. However, close to 60\% of the data belongs to individuals from the following countries: the United States, United Kingdom, Germany, India, and France. Our work here extends on Voxceleb's database by including primarily three nationalities: Chinese, Japanese, and Korean.


The work here primarily makes two contributions to speaker recognition in ``real-world" situations. First, inspired by the Voxceleb dataset acquisition method, we develop a fully automated and scalable pipeline that can collect audio-visual recordings of famous individuals. The developed system takes as input the name of an individual and returns as output segments within various Youtube videos where that person was talking. For any arbitrary set of speakers from any arbitrary language, our system can collect data from that language. Collecting a language-specific dataset is of utmost importance, considering it can tune speaker recognition models for specific languages. Even though the system pipeline follows the overall procedure proposed by Voxceleb, the implementation details are very different for the various subsystems involved in the architecture. Subsequent sections will discuss this system in detail. 

The second contribution of the work shows that applying diarization on obtained Youtube audio can decrease the error rate. Other methods have used face-tracking (which can accumulate error over time) or frequent face detection (which is much slower) to identify segments in which the desired talker is present. To mitigate the problems with face-tracking, the diarization of audio recordings can be an effective way to improve the overall accuracy of the dataset. Furthermore, by applying diarization, we can remove undesired audio obtained during our pipeline. On average, diarization led to a 9.7\% improvement in equal error rate during our verification experiments.

Overall, we collected unique voice recordings for 1,900 celebrities across three countries. We set up a verification system to evaluate the validity and accuracy of the collected data. Our goal was to show that our proposed method can effectively collect speaker data with a relatively low error rate. This error rate further decreased after the applied diarization. EACeleb can be used both for speaker verification and identification tasks. We use the x-vector system implementation in Kaldi toolkit to train and test our data.

\section{Data Collection Pipeline}
\label{sec:dataCollection}

The data collection pipeline follows the general ideas proposed by Voxceleb. However, the implementations of the different components of the pipeline are unique. In many ways, our work is optimized for speed, particularly during the face-matching phase of the pipeline. Nevertheless, our system generalizes to celebrities of different countries and even people who are not celebrities but have some Youtube presence.

\begin{table}[h!]
    \centering
    \begin{tabular}{| c | c |}
         \hline
         \textbf{\# of Chinese POI} &  $410$ \\
         \hline
         \textbf{\# of Japanese POI} & $406$ \\ 
         \hline
         \textbf{\# of South Korean POI} & $925$ \\
         \hline
         \textbf{Average \# of Utterances per POI} & $46.6$ \\
         \hline
         \textbf{Average Duration per Utterance} & $8.4~sec$ \\
         \hline
         \textbf{\# total of utterances}  & $76522$ \\
         \hline
         \textbf{\# total audio hours}  & $180$ \\
         \hline
         \hline
    \end{tabular}
    \caption{Dataset Statistics for EACeleb (POI - Person of Interest)}
    \label{tab:ratio_data}
\end{table}

\subsection{Template Face}
The first step in our pipeline is the creation of a template face. Here, we automatically use a Google search to download the top 20 search results of the person of interest. Then, we append the word "photo" or "face" to the person's name before the search. For example, suppose the celebrity's name is ``John Doe". The search term for our system will be ``John Doe Face." We append this term so our output images will more likely be face shots rather than full-body captures. The face shots are essential for creating the template face.

Note that each resulting image could have multiple faces present in the image. For each photo, we use a histogram of gradients (HOG)\cite{hog} based face detection to find all faces in the picture. Next, we use VGG-face to extract a 2048-dimensional vector for all faces found. Last, we use a density-based spatial clustering of applications with noise DBSCAN \cite{dbscan} to cluster the extracted embeddings. We heuristically identify the largest cluster as the face of the celebrity of interest. We take the average of all the embeddings in the largest cluster as the template face of the celebrity. If the number of embeddings in the largest cluster is less than 10, we disregard that celebrity.

\subsection{Video Downloading}
The next task is downloading videos from Youtube of the celebrity. Like Voxceleb, we append the word "interview" to the celebrity's name before the Youtube search. Then, we download the top 15 results of this Youtube search. For both the Google images and Youtube video download, we searched the names written in their appropriate language of interest. We also translated the words "face," "photo," and "interview" to the languages matching the celebrity's nationality.

\subsection{Video Processing}
Like Voxceleb, the first step for processing the Youtube video is applying shot boundary detection. This is accomplished using the pyscenedetect library, which uses a color histogram for shot boundary detection. Subsequent steps in our video processing, however, differ from Voxceleb. For Voxceleb's implementation, face detections are grouped into face-tracks/tracklets within each detected shot. Their implementation followed the pipeline described in \cite{lip_reading}\cite{out_of_time}. Essentially, every face that appears in the video has a corresponding tracklet. Using a tracklet is expensive, considering it requires face detection in every frame and storing tracklets for all faces. Our implementation, however, combines detection and verification to build tracklets for only the person of interest.

\subsection{Face Verification}
Instead of tracking all faces in each detected shot, we use a HOG-based procedure to find faces in a frame and compare each of the faces with the template face created earlier using the Google downloaded images. Once we find the face of the person-of-interest (POI), a CSRT-based algorithm \cite{csrt} along with face pixel histogram is used to track the POI in subsequent frames. This procedure is more efficient than the re-detection of faces in successive frames. Additionally, it eliminates the need to create tracklets and determine which tracklets belong to the POI.

\subsection{Speaking Detection}
Once we have collected tracks from the person-of-interest, the next phase is determining segments in the tracklet where the person-of-interest POI is actually talking. We applied the Syncnet tool developed by \cite{syncnet} here. Syncet was originally designed to determine the audio-video synchronization between mouth motion and produced speech. It can, however, be repurposed to determine whether or not a POI is talking in a particular segment.

\begin{table*}[!htbp]
    \centering
    \begin{tabular}{| c | c | c|c|c|c|c|c| c | c |}
         \hline
         \textbf{Model} & \multicolumn{3}{c|}{\textbf{Voxceleb}} & \multicolumn{3}{c|}{\textbf{SITW}} & \multicolumn{3}{c|}{\textbf{SRE16}}\\
         \hline
         & VSR & EER\% & MinDCF& VSR & EER\% & MinDCF& VSR & EER\% & MinDCF\\
         \cline{2-10}
         70 & 0.54 & 3.14 & 0.35 & 0.52 & 3.25 & 0.37 & 0.39 & 3.18 & 0.22 \\
         80 & 0.79 & 6.83 & 0.52 & 0.72 & 5.55 & 0.46 & 0.57 & 5.44 & 0.39 \\
         100 & 0.99 & 15.09 & 0.66 & 0.95 & 15.23 & 0.68 & 0.87 & 15.8 & 0.66\\
         \hline
    \end{tabular}
    \caption{Useful Collection Ratio (UCR) of collected speakers (SRE diarization, in percentage). \\ 
    $VSR = \frac{Number of segments ~kept ~per ~speaker ~after~ diarization}{total ~number ~of ~speaker~ segments~ before ~diarization}$}
    \label{tab:ratio_data}
\end{table*}

\section{Experiments}
\label{sec:majhead}

We used a speaker verification task on the acquired data to verify the validity of the data collection procedure. The primary metric to evaluate performance is the equal error rate (EER). One particular area of focus is the role of diarization for further cleaning the collected data. 

For our experiments, we use pre-trained x-vector models with a configuration similar to the one described by Snyder et al. \cite{xvectors}. An x-vector system is a feed-forward DNN that maps a variable-length speech segment to a fixed-length embedding. The input to the system is 24-dimensional MFCC features extracted every 10 milliseconds.

The first five levels of this DNN operate at the frame level with a temporal context centered at current frame t. For example, frame1 has a full context of 5 since it concatenates features from t - 2, t - 1, t, t + 1, and t + 2. frame2 builds on frame1 by splicing outputs from frame1 at t-2, t, and t+2. This gives frame2 a total temporal context of 9. The process goes on for the five frame-level layers. Mean and standard deviation values are calculated in the statistics pooling layer. This layer aggregates all T frame-level outputs from frame5. X-vectors are 512-dimensional embeddings extracted at segment6.

The x-vector network is implemented in the nnet3 neural network library in the Kaldi \cite{kaldi} Speech Recognition toolkit.

\subsection{Experiments Using Pretrained Speaker Models}
\label{ssec:subhead}

The first series of experiments evaluated the EACeleb dataset accuracy. To accomplish this, we set up a speaker verification task similar to the one described in \cite{xvectors}. Utterance durations range from 4 seconds to 20 seconds. We pass each utterance through a pre-trained network to extract the x-vectors. We used models pretrained with the SITW, Voxceleb, and SRE16 datasets for experimentation. After training, we then extract 512-dimensional x-vectors from segment 6. 

We split the extracted x-vectors into a training and testing set. 80\% of the speakers were reserved for training. The remaining speakers comprised our test set. Using the training set, we trained both a linear discriminant analysis (LDA) matrix and a probabilistic LDA (PLDA) model. The test set was used for speaker verification. A ``test x-vector" belonging to a particular speaker is compared to all the other x-vectors from that same speaker and given a ``target" label. Equivalently, we sampled an equal number of x-vectors coming from speakers different from the "test x-vector" speaker. We compare the sampled x-vectors with the test x-vector and assign a ``non-target" label. 
\subsection{Diarization Experiments}
\label{ssec:subhead}
We applied diarization on a per speaker basis. Specifically, x-vectors belonging to a particular speaker were clustered using a DBSCAN algorithm similar to how we clustered faces in our data collection pipeline. For each speaker, we first use PLDA scores to compare all combinations of x-vectors from that speaker with each other. We then use the PLDA scores as the distance metric values for our DBSCAN clustering algorithm. The underlying assumption is that the largest cluster from the DBSCAN will belong to the speaker. Smaller clusters will discarded away as inaccurate speaker data. DBSCAN requires a parameter, eps, the maximum distance between two x-vectors for one x-vector to be considered in the neighborhood of another x-vector. We experimented with tuning this eps parameter.
 
After this data cleaning method, we use the same train-test split as in pre-diarization verification. We set up a speaker verification task similar to the setup in the previous section and calculated the error rates.

 \subsection{Fine-Tuning Experiments}
\label{ssec:subhead}
 For the last series of experiments, we employed an active learning technique to improve both the model we are using to extract the x-vectors and clean the dataset further. We wanted to fine-tune the x-vector model for the EACeleb dataset. We started by first applying our diarization technique to both the train and test set to accomplish this. Audio identified by the diarization as belonging to a particular speaker became the primary data we used. We applied data augmentation to this raw audio. This includes adding MUSAN music, noises, and reverberation to each audio utterance. We follow the same augmentation technique outlined in the kaldi voxceleb recipe. Next, we reconfigured the last layer of the x-vector to have the same number of nodes as the number of speakers in our train set. Accordingly, this reinitializes the weights of the final layer which we trained. We then retrained the x-vector model
 
 After retraining the model, we extracted x-vectors for the train and test data and set up another verification task. Then, LDA and PLDA models were again trained using the train data.

\begin{table}[!htbp]
    \centering
    \scalebox{0.9}{
    \begin{tabular}{| c | c | c | c |}
         \hline
         \textbf{Pretrained Model} & \textbf{VoxCeleb} & \textbf{SITW} & \textbf{SRE}\\
         \hline
             \begin{tabular}{ c  c }EACeleb EER \\ before Diarization
    \end{tabular} & 15.09  & 15.23& 15.83 \\
         \hline
       \begin{tabular}{ c  c }EACeleb EER \\ After Diarization (eps=80)
    \end{tabular} & 6.833 & 5.55  & 5.42 \\ 
         \hline
         \begin{tabular}{ c  c }EACeleb EER on Fine-Tuned\\ X-vector(eps=80) 
    \end{tabular} & 3.833 & 4.86  & 5.31 \\ 
         \hline
    \end{tabular}}
    \caption{EER\% performances of our trained models (in percentage).}
    \label{tab:eer_train}
\end{table}

    \begin{tabular}{| c | c }
    \end{tabular}

\section{Results and Discussions}
\label{sec:print}

For comparison, we selected ten videos of approximately 1 hour each and compared the speed to complete the audiovisual search using face-detection only versus the case of coupling face-detection with tracking. For those ten videos, using CSRT tracking once a face has been found was about eight times faster. Making this kind of comparison is difficult since many factors can affect the speed of the tracking method.

Despite the improved speed with our face-tracking method, we saw significantly higher error rates when compared to the Voxceleb dataset. Table \ref{tab:eer_train} summarizes the error rates using the raw audio collected from Youtube for our EACeleb dataset. On average, the error rate was about 15\% . In comparison, the Voxceleb dataset on the x-vector model had an error rate of only 3.1\%. Our high error rate can be attributed to our face tracking algorithm not being as accurate as constantly doing face detection for subsequent frames. When there is a scene change, the tracking algorithm is not robust and tracks a person, not of interest and stores audio for that person. By applying diarization, we decreased the error rate to approximately 6\%.

We experimented with various eps values used in DBSCAN for our diarization. The values of eps we chose were chosen from 70, 80, and 100. Over all eps range, x-vectors extracted from Voxceleb based models were more robust. Table \ref{tab:ratio_data} shows the performance for various eps. DBSCAN essentially created one huge cluster for high value of eps, signifying that all the x-vectors belong to the speaker. This increased the equal error rate. The goal is to choose eps to reduce equal rates without totally discarding true speaker segments.

SRE16 performed the worst among all the pretrained models. DBSCAN was not able to cluster x-vectors extracted with the SRE16 model. This is because SRE16 was trained with clean data. For low values of eps, SRE16 based models did not correctly cluster segments belonging to a speaker. This explains low VSR rates. Even though it led to decreased error rate, it's not preferable. Voxceleb, on the other hand, achieved error of 6.833\% even after keeping nearly 80\% of data post diarization. Applying the fine-tuning technique to the Voxceleb data decreased our error rate to about 3.8\%. The resulting model is more adapted to our dataset.

\section{Conclusions}
\label{sec:page}
In this paper, we describe a pipeline that can effectively be used to build a speakers database. Here, our database consisted of speakers primarily in East Asian countries. We also show that applying diarization on collected data can be an effective method for further cleaning the Youtube data. Our EACeleb dataset of 1,641 speakers can be used for speaker verification and identification tasks. For individuals in the speech research community who want to tune their models for speakers of a specific language, we hope the pipeline can serve as a guide for acquiring that language-specific data. The fine-tuning experiment showed that this data could further be used to tune pre-trained models.

\vfill\pagebreak

\bibliographystyle{IEEEbib}
\bibliography{mybib}

\end{document}